# Integrated quantized electronics: a semiconductor quantized voltage source

**Authors:**
Frank Hohls, Armin C. Welker, Christoph Leicht, Lukas Fricke, Bernd Kaestner, Philipp Mirovsky, André Müller, Klaus Pierz, Uwe Siegner, and Hans Werner Schumacher

**Affiliation**
Physikalisch-Technische Bundesanstalt, Bundesallee 100, 38116 Braunschweig, Germany

Corresponding Author:  Hans Werner Schumacher
E-Mail:               hans.w.schumacher@ptb.de
Phone: +49 531 592 2500   Fax:    +49 531 592 69 2500

**Abstract:**

The Josephson effect in superconductors links a quantized output voltage $V_{\text{out}} = f \cdot (h/2e)$ to the natural constants of the electron's charge $e$, Planck's constant $h$, and to an excitation frequency $f$ with important applications in electrical quantum metrology. Also semiconductors are routinely applied in electrical quantum metrology making use of the quantum Hall effect. However, despite their broad range of further applications e.g. in integrated circuits, quantized voltage generation by a semiconductor device has never been obtained. Here we report a semiconductor quantized voltage source generating quantized voltages $V_{\text{out}} = f \cdot (h/e)$. It is based on an integrated quantized circuit of a single electron pump operated at pumping frequency $f$ and a quantum Hall device monolithically integrated in series. The output voltages of several µV are expected to be scalable by orders of magnitude using present technology. The device might open a new route towards the closure of the quantum metrological triangle. Furthermore it represents a universal electrical quantum reference allowing to generate quantized values of the three most relevant electrical units of voltage, current, and resistance based on fundamental constants using a single device.

**Article:**

The transition from multiple interconnected electronic components to integrated circuits revolutionized the electronics industry and enabled highly integrated devices with new and complex functionalities [1]. With increasing integration and reduced structural dimensions quantum size effects become more and more important and will eventually strongly affect device operation. One *prospect* of these quantum size effects is the quantization of electrical transport quantities, which involves two natural constants: the charge of the electron $e$ and Planck's constant $h$. Making use of this quantization in *quantized* electronic devices (see e.g. [2,3]) might enable new device functionalities based on their high absolute precision and universality. Among the quantization effects occurring in semiconductor devices are e.g. quantum-Hall resistance quantization in units of $h/e^2$ [4], one dimensional conductance quantization in units of $e^2/h$ [5,6], single charge quantization [7], as well as single charge based current quantization in units of $ef$ where $f$ is the frequency of a suitable AC drive voltage [8]. In superconducting Josephson junctions [9] also quantized voltages in multiples of $f \cdot (h/2e)$ can be generated with important applications in electrical quantum metrology [10]. However, in semiconductor devices such generation of quantized voltages has not been possible up to now.

Here we report on an all-semiconductor quantized voltage source allowing the generation of quantized voltages $V_{\text{out}} = f \cdot (h/e)$ upon application of a suitable AC drive voltage signal with frequency $f$. The device is based on an integrated *quantized* circuit (IQC) consisting of a non-adiabatic single-electron pump [11] and a quantum-Hall device monolithically integrated in series. Robust operation with output voltages of several µV corresponding to operating frequencies of a few GHz is demonstrated. The output voltage is expected to be scalable by orders of magnitude using present technology. By using either the functionality of the complete IQC or the functionalities of its two components it could further serve as a *universal* electrical quantum reference allowing to link the three most relevant electrical units of voltage, current and resistance to the fundamental constants $e$ and $h$ in a single device.

The quantum Hall effect in two-dimensional semiconductors quantizes the relation between the applied current $I$ and the generated Hall voltage $V_H$ by

$$V_H = I \frac{1}{\nu} \frac{h}{e^2}, \qquad (1)$$

where the so called filling factor $\nu$ is an integer determined by the applied magnetic field and the carrier density of the Hall bar. Its observation and precision quantization in different material systems like silicon [4], gallium arsenide [12], and more recently in graphene [13,14,15] provides one of the most stringent tests of the universality of this relation. Likewise, generation of quantized currents of amplitude

$$I = nef \qquad (2)$$

has been obtained in different metallic [16,17,18,19] and semiconducting material systems [11,20,21,22] by periodically shuttling a number of $n$ electrons from source to drain. Up to now such components have generally been viewed as separate entities. However, by using devices based on the same material system like GaAs their integration into an IQC becomes a straightforward option to realize new device functionalities.

The IQC under investigation consists of a recently developed high current non-adiabatic quantized charge pump [11] and a quantum Hall resistor connected in series, as illustrated in Fig. 1(a). Both devices are fabricated from a two-dimensional electron system based on a high mobility modulation doped GaAs/AlGaAs heterostructure (see methods).

Fig. 1(b) shows a scanning electron micrograph of a typical non-adiabatic single electron pump. The pump consists of a 900 nm wide etched constriction in the GaAs/AlGaAs heterostructure crossed by three narrow metallic top gates $G_1$ - $G_3$. $G_3$ is not essential for pumping and is generally grounded. Negative DC voltages $V_1, V_2$ are applied to $G_1$ and $G_2$, respectively, to define two electrostatic barriers and a quantum dot (QD) in between. To induce pumping an additional sinusoidal AC gate voltage component $V_{AC}$ with drive frequency $f$ and AC power $P_{AC}$ is applied to $G_1$. The principle of the resulting pumping cycle is sketched schematically in Fig. 1(c). During the first half-cycle of the AC oscillation the left barrier to the source lead oscillates down and becomes transparent. At the same time the bound state of the QD drops below the chemical potential µ of the source and the QD is loaded with electrons from the source reservoir. This phase of the pumping cycle is sketched by the blue potential landscape in the figure. In the following phase of the pumping cycle the height of the source barrier raises and the barrier becomes more opaque. At the same time the QD ground state raises above µ and the QD becomes more confined. During this initialization phase the excess electrons that were captured on the QD during the first phase of the cycle are ejected back to source resulting in a high fidelity initialization of the dynamic QD with $n$ electrons [23]. In the final phase of the pumping cycle (sketched orange) the $n$ electrons are unloaded to the drain reservoir (right hand side) which is connected to the Hall bar before the cycle starts over again. Under continuous AC excitation the device then acts as a quantized current source generating a current of $I = nef$, where the number $n$ of pumped charges can be controlled by the DC gate voltages $V_1$ and $V_2$. A more detailed description of the physics underlying non-adiabatic single electron pumping can be found e.g. in Refs. 11 and 23.

Fig 1(d) shows a typical operation curve of such non-adiabatic single electron pump. The pumped current is plotted as function of $V_2$ for fixed $V_1$ and $P_{AC}$. Two quantized current plateaus corresponding to quantized currents of $I = nef$, $n = 1,2$ are well observed. With increasing $V_2$ (i.e. towards the right hand side in Fig. 1(d)) the height of the drain barrier decreases. At the same time both the confinement of the QD and the energy of the lowest bound QD state decrease. Hence the number $n$ of electrons that are kept in the dot during the initialization phase increases stepwise with $V_2$ leading to the observed stepwise increase of $I$. Note that precision measurements of the quantized currents of similar devices reported by Giblin et al [24] have confirmed current quantization within a measurement uncertainties of 15 ppm. Note further that the optimum expected uncertainty of the current quantization can be theoretically evaluated by fitting the measured $I(V_2)$ curves to the so-called decay cascade model for quantum dot initialization [23]. Doing so a minimum quantized current uncertainty of about $10^{-8}$ can be estimated from the data of Fig 1(d). It should be stressed that the current quantization of these non-adiabatic single electron pumps is robust against variations of the applied DC and AC gate voltages, against an applied bias voltage [25], and against applied magnetic fields up to several tesla [26]. The latter two properties are very important for the IQC under investigation. While the bias robustness enables on-chip integration with components representing an ohmic load for the pumped current like in our case the Hall device, the magnetic field robustness enables quantized current generation while the Hall device is operated in the quantum Hall regime.

Fig. 1(e) shows the Hall resistance $R_{xy}$ as function of the applied perpendicular field $B$. $R_{xy}$ reveals the well known Hall quantization with quantization plateaus of $R_{xy} = \nu^{-1} \cdot (h/e^2)$ at integer filling factors $\nu = 1,2,3,…$ The two most robust resistance plateaus with $\nu = 1$ and $\nu = 2$ are situated at magnetic fields around 8.4 T and 4.2 T, respectively, defining the optimum operation fields for the semiconductor quantized voltage source.

When operated at these fields the IQC consists of a single charge pump delivering a quantized current according to Eq. (2) and a serially connected quantum Hall device converting the current into a quantized voltage according to Eq. (1). The quantized output voltage should hence be given by

$$V_{out} = \frac{n}{\nu} \frac{h}{e} f \ . \tag{3}$$

$V_{out}$ should thus be determined by the two natural constants *e* and *h*, by the two integers *n* and ν, and by the applied drive frequency *f*. It should thus reveal a quantization relation similar to the Josephson relation, however, of course based on completely different physics.

Fig. 2 shows typical measurements of quantized voltage generation of two of our IQC devices. In (a) the device already characterized as current source in Fig. 1(d) is now operated as voltage source at $f = 675$ MHz and at B = 8.4 T corresponding to filling factor ν = 1. The output voltage $V_{out}$ is plotted as function of $V_2$. Similar to the case of the current quantization (cf. Fig. 1(d)) now three quantized *voltage* plateaus corresponding to quantized voltages of $V_{out} = n\frac{h}{e}f$ with $n = 1,2,3$ are found in the displayed data range. Quantized voltage plateaus are also observed as function of both DC voltages $V_1$ and $V_2$ as shown in Fig. 2(b). Here, $V_{out}$ is displayed in a three dimensional plot as function of $V_1$ and $V_2$. The plot reveals a well defined voltage plateau (green) at $n = 1$. The generated quantized voltage is thus robust against variation of $V_1$ and $V_2$. The shape of the plateau as function of $V_1$, $V_2$ is again determined by the shape of the quantized current plateau of the non-adiabatic pump [27].

A test of the robustness of the voltage quantization against variation of the other two relevant external parameters *B* and $V_{AC}$ is displayed in Fig. 2(c) and (d), respectively. To test the influence of a variation of *B within* the quantum Hall plateaus at ν = 1,2 the average $V_{out}$ within the $V_{out}(V_2)$ plateau and its statistical uncertainty is determined for different values of *B*. In (c) $V_{out}$ is plotted for $f = 320$ MHz in natural constants as function of magnetic fields around ν = 1 (red dots, upper scale) and ν = 2 (blue dots, lower scale). For the whole field range and both filling factors $V_{out}$ remains well quantized within the statistical uncertainty as indicated by the error bars (cp. methods). Next, the dependence of the voltage quantization on the amplitude $V_{AC}$ and hence on the applied AC excitation power $P_{AC}$ will be considered. The non-adiabatic pump requires a minimum threshold input power beyond which the generated current remains fixed at its quantized value [27]. This robustness against variation of $P_{AC}$ is well transferred to the operation of the IQC as shown in Fig. 2(d). Here, $V_{out}$ is plotted as function of $P_{AC}$ for excitation powers of -16 … -10 dBm. The corresponding AC voltage amplitude $V_{AC}$ at $G_1$ can be estimated to be in the range of 100 … 200 mV [27]. Parameters for the given data are $f = 1.1$ GHz, $\nu = 1$, $n = 1$ (black squares) and $n = 2$ (orange dots). Again, no significant dependence of $V_{out}$ on $P_{AC}$ is observed. Note that for the displayed data the statistical uncertainty is smaller than the dot size in the graph. Hence, $V_{out}$ can be considered to be robust over a broad range of applied AC voltages.

Finally the linear dependence of $V_{out}$ with $f$ as predicted by Eq. (3) is tested. Here, the output voltage $V_{out}$ inside the plateaus is plotted as function of $f$. Fig. 3 compiles data measured on two devices using two different measurement setups (cp. methods) during a number of cool down cycles. To collect the data the devices were operated inside the robust operation region of the control parameters $V_1$, $V_{AC}$, and $B$. Then traces of $V_{out}(V_2)$ (cp. Fig. 2(a)) were taken. The average $V_{out}$ in the plateau and its statistical uncertainty was derived. Three groups of data sets are found in the plot. They can be classified according to the two integers $n$ and $\nu$ and hence according to the operation parameters of the two devices forming the IQC. The black points correspond to $V_{out}$ measured for $n = \nu = 1$, the blue points correspond to $n = 1$, $\nu = 2$, and the red ones correspond to $n = 2$, $\nu = 1$. All data sets are well described by the theoretically expected curve according to Eq. 3 as indicated by the three dashed straight lines. The data clearly evidences the expected quantization of the generated voltage in units of $h/e$. To evaluate quantitatively the agreement with the theoretical prediction we fit the slope of the measured voltages of Fig. 3 for $n = \nu = 1$. The result of 4.1330(12) µV/GHz deviates less than 1‰ from the expected value of 4.1356 µV/GHz. Note that this deviation is well within the systematic uncertainties given by our measurement setups (see methods) and hence no significant deviation of $V_{out}$ from the theoretically predicted value is evident. The data presented in Fig. 3 clearly shows that the IQC is capable of generating quantized voltages above 10 µV corresponding to operation frequencies of about 3 GHz. Higher output voltages can be expected for optimized devices and optimized experimental conditions. Furthermore $V_{out}$ is expected to scale linear with the number of the integrated devices of the IQC. For example the pumped current could be increased by operation of a number of $N_P$ single-electron pumps in parallel [28, 29, 30] thereby also scaling up the output voltage by $N_P$. Likewise, instead of a single Hall device a serial array of $N_H$ quantum Hall devices could be used [31] resulting in an accordingly scaled output voltage. Combining both concepts and employing values for the available technology of $N_P \sim 10$ [28,30] and of $N_H \sim 100$ [31] scaled up quantized output voltages of the order of tens of millivolt seem to be realistic.

The above data illustrates that our semiconductor IQCs indeed fulfills clearly the new semiconductor device functionality of robust quantized voltage generation. The IQC thus links the output voltage to the two fundamental natural constants $h$ and $e$ with high precision and takes advantage of their universality. We anticipate the three main features of robustness, universality, and precision to be the major new elements in the functionality of future IQCs, reflected in the discrete dependence of the electrical output quantity on the input. The universality of this IQC device concept should in principle allow for its transfer to different material systems like silicon [21,4] or even graphene [13-15] provided a suitable graphene based quantized charge pump can be realized. One natural area of application of the given IQC is the field of electrical quantum metrology, which aims at defining the units of physical quantities using true invariants of nature [32]. Here the presented IQC might allow for an alternative approach to the closure of the quantum metrological triangle [33] by direct high precision comparison of the (scaled up) output voltage of a semiconductor quantized voltage source to the output voltage of a Josephson based quantum voltage standard [10]. Furthermore it can be viewed as a *universal* reference for electrical quantum metrology as it allows the generation of quantized values of the three most relevant electrical units volt, ampere, and ohm based on the fundamental constants *e* and *h* using a single device.

**Acknowledgement**

This work has been supported by the DFG, by EURAMET joint research project with European Community's Seventh Framework Programme, ERANET Plus under Grant Agreement No. 217257, and by the Centre for Quantum Engineering and Space Time research QUEST. C.L. has been supported by International Graduate School of Metrology, Braunschweig.

**Authors contributions**

C.L., B.K., P.M., and K.P. contributed to sample fabrication. A.C.W., C.L., F.H., L.F. and B.K. performed the experiments. F.H., A.C.W., C.L., L.F., A.M., B.K., U.S. and H.W.S. contributed to the analysis and discussion of the experiments. F.H., A.C.W., C.L. B.K., U.S. and H.W.S. contributed to writing the manuscript, and H.W.S. initiated and supervised the research.

**Methods:**

Devices are fabricated from a GaAs/AlGaAs based modulation doped heterostructure containing a two-dimensional electron system situated 95 nm below the surface with carrier concentration $N_e = 2.1 \times 10^{15}$ m$^{-2}$ and electron mobility of 97 m$^2$/Vs at 4.2 K. The conducting semiconductor channel of the pump and the Hall bar are defined simultaneously by electron-beam lithography and wet-chemical etching. Ohmic contacts are defined by alloyed AuGeNi. TiAu op gates are defined by electron-beam lithography and lift off. The gates have a width of 100 nm and 250 nm centre spacing. Transport measurements were done in a ³He cryostat at base temperature of $T = 350$ mK. Perpendicular magnetic fields up to $B = 15$ T were applied by a superconducting coil. Two devices with identical layout were measured using two different measurement setups. To suppress drifting offset voltages of the measurement circuits differential measurements were carried out by comparing the output signal in the region of a pumped current and in the region of quenched pumping of the single electron pump.

The voltage generated by device **A** was measured using a calibrated HP 34420 nano voltmeter. Here the drifting offset voltage was determined before each gate voltage variation for quenched pump action at sufficiently negative gate voltage $V_2$ and subtracted from the measured curves. The quantized $V_{out}$ were determined by averaging multiple values on the flat part of the plateau, with the uncertainty derived from the variance of the measured voltage. Device **B** was measured in a switched current setup. Here $V_2$ was switched multiple times between a sufficiently negative gate voltage for quenched pump action (OFF) and the chosen working point (ON). After each switching the voltage was amplified by a Femto DLPVA voltage amplifier with 1000 ($\pm 1\%$) fold gain and 1 G$\Omega$ input impedance and measured by an analog-digital converter. The difference between ON and OFF pumping state for each switching is averaged over a total of 200 to 800 switching actions with 5 to 12 gate voltage changes per second. This procedure allows for long averaging of the generated voltage without suffering from strong 1/f noise and drifts. The typical remaining statistical uncertainty due to noise is $< 0.005$ µV. The dominating error source in the measurement of device **B** is the systematic gain uncertainty of the voltage amplifier, specified with 1%. The linear least square fit of the frequency dependence of Fig. 3 was performed for fixed zero offset and with data weighting by the statistical uncertainty.

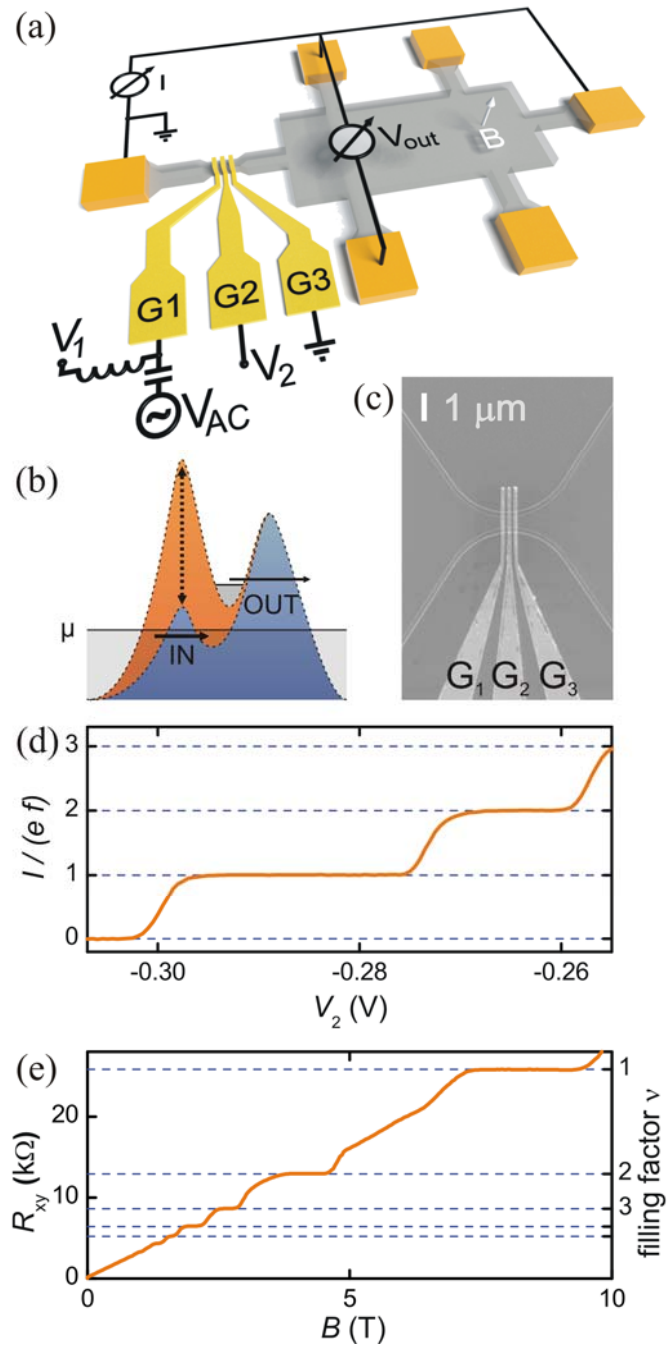

**Fig. 1:** (a) Schematic picture of the device and measurement setup: A single electron pump (left) is connected in series with a Hall bar (right). Gray: semiconductor with two dimensional electron system; gold: Ohmic contacts; yellow: top gates (b) Potential-landscape of the pump at the loading and unloading phase, respectively. (c) SEM picture of typical pump. (d) Pumped current $I$ vs. $V_2$ at $f = 675$ MHz. $V_1 = -0.325$ V, $P_{AC} = -16$ dBm, $B = 4.2$ T, $T = 350$ mK. (e) Quantum Hall effect of the Hall bar for a measurement current of 1 µA, $T = 350$ mK.

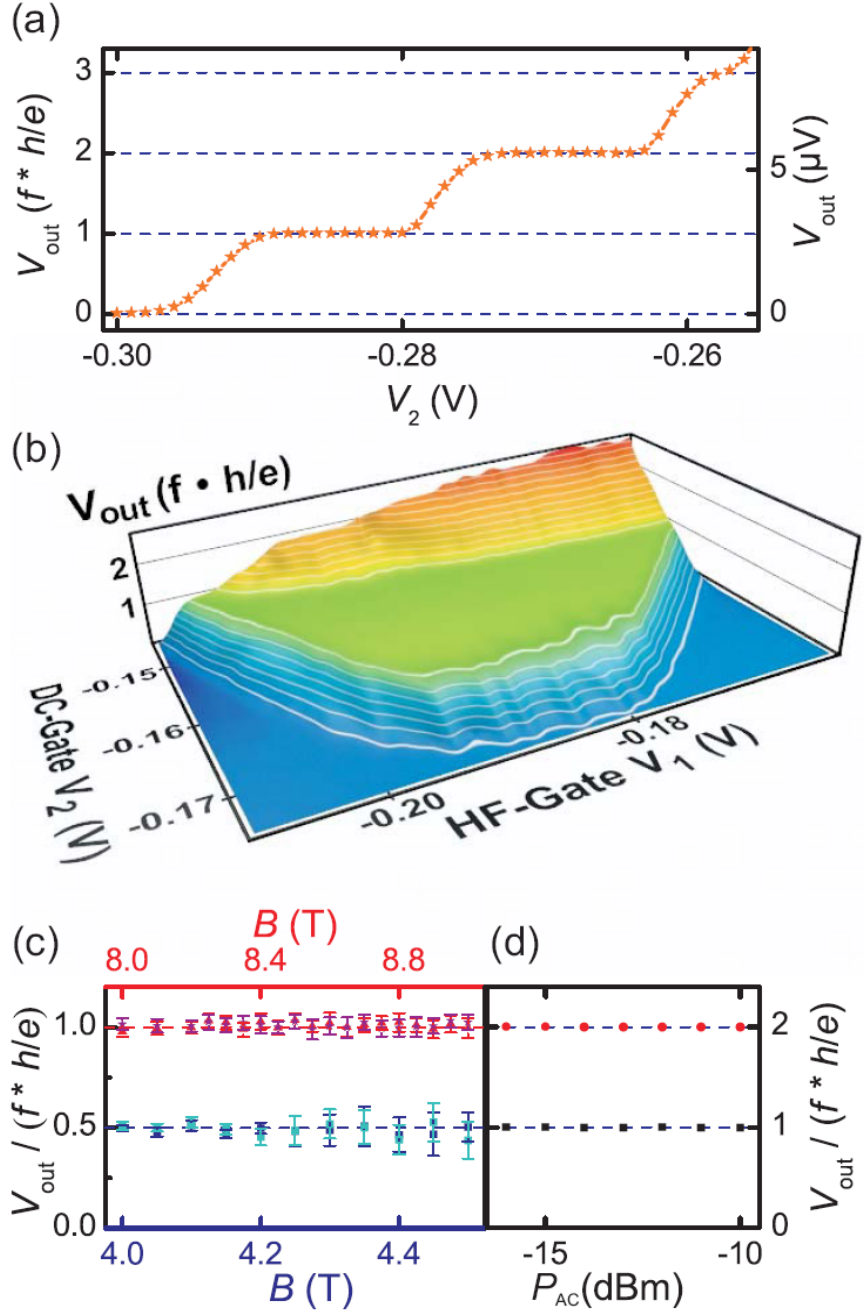

**Fig. 2:** Quantum voltage source operation. (a) Output voltage $V_{out}$ as function of $V_2$. (b) Voltage plateau as function of $V_1$ and $V_2$. $\nu=1$ and $f = 675$ MHz in (a) and (b). Data in (a),(b) is taken in different cool down cycles resulting in a shift of the absolute values of $V_1$, $V_2$ at which the plateaus are observed. (c) Robustness of $V_{out}$ for $\nu=1$ and $\nu=2$ against variations of $B$ within the resistance plateaus. $f = 320$ MHz and $P_{AC} = -6$ dBm (device A). The different colours correspond to different magnetic sweep directions. (d) Robustness against variations in excitation power at $f = 1.1$ GHz , $\nu = 1$, $n = 1$ (black squares) and $n = 2$ (orange points). Data in (a,b,d) is taken on device B, data in (c) is taken on device A. Error bars correspond to the statistical measurement uncertainty. If not plotted the error bars are smaller than the symbol size.

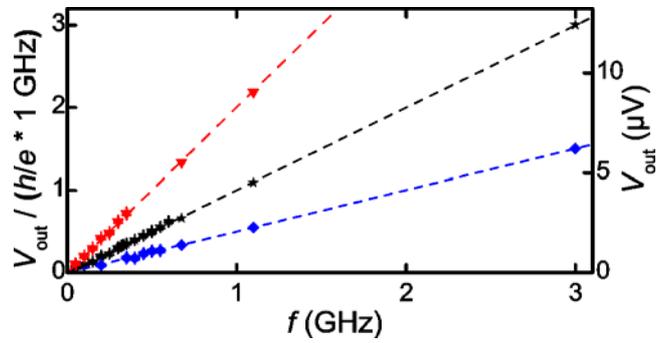

**Fig. 3:** Frequency dependence of the generated plateau voltage $V_{out}$. The data points are obtained from averaging over $V_2$ within the *nef*-plateau region. Data is collected for different cool down cycles and different $V_1$, $P_{AC}$ parameters in the centre of the quantum Hall plateaus on two devices (A,B). The three data sets with different slopes correspond to $n = 1$, $\nu = 1$ (black), $n = 2$, $\nu = 1$ (red) and $n = 1$, $\nu = 2$ (blue). The three dashed lines show the expected values according to Eq. 3. Symbols correspond to measured data. Statistical measurement uncertainty is smaller than the symbol size.